\documentclass[default,iicol]{sn-jnl}

\usepackage{lineno}

\jyear{2021}%

\theoremstyle{thmstyleone}%
\theoremstyle{thmstyletwo}%

\theoremstyle{thmstylethree}%

\usepackage{subfig}

\begin{document}





\title[Article Title]{GNN for Deep Full Event Interpretation  and hierarchical reconstruction of heavy-hadron decays in proton-proton collisions}


\author*[1,2]{\fnm{Juli\'an} \sur{Garc\'ia Pardi\~nas}}\email{julian.garcia.pardinas@cern.ch}

\author[1]{\fnm{Marta} \sur{Calvi}}\email{marta.calvi@cern.ch}

\author[3]{\fnm{Jonas} \sur{Eschle}}\email{jonas.eschle@cern.ch}

\author[4,5]{\fnm{Andrea} \sur{Mauri}}\email{a.mauri@cern.ch}

\author[1]{\fnm{Simone} \sur{Meloni}}\email{sim.meloni@gmail.com}

\author[1]{\fnm{Martina} \sur{Mozzanica}}\email{martina.mozzanica@cern.ch}

\author[3]{\fnm{Nicola} \sur{Serra}}\email{nicola.serra@cern.ch}

\affil[1]{\orgdiv{Dipartimento di Fisica ``G. Occhialini''}, \orgname{Universit\`a di Milano Bicocca and INFN Sezione di Milano-Bicocca}, \orgaddress{\street{Piazza della Scienza 3}, \city{Milano}, \postcode{20126}, \country{Italy}}}

\affil[2]{\orgdiv{Experimental Physics Department}, \orgname{European Organization for Nuclear Research (CERN)}, \orgaddress{\street{Espl. des Particules 1}, \city{Meyrin}, \postcode{1211}, \country{Switzerland}}}

\affil[3]{\orgdiv{Department of Physics}, \orgname{University of Z\"urich}, \orgaddress{\street{Winterthurerstrasse 190}, \city{Z\"urich}, \postcode{8057}, \country{Switzerland}}}

\affil[4]{\orgname{Nikhef National Institute for Subatomic Physics}, \orgaddress{\street{Science Park 105}, \city{Amsterdam}, \postcode{1098 XG}, \country{The Netherlands}}}

\affil[5]{\orgname{Imperial College London}, \orgaddress{\street{South Kensington Campus}, \city{London}, \postcode{SW7 2AZ}, \country{UK}}}


\abstract{The LHCb experiment at the Large Hadron Collider (LHC) is designed to perform high-precision measurements of heavy-hadron decays, which requires the collection of large data samples and a good understanding and suppression of multiple background sources. Both factors are challenged by a five-fold increase in the average number of proton-proton collisions per bunch crossing, corresponding to a change in the detector operation conditions for the LHCb Upgrade I phase, recently started. A further ten-fold increase is expected in the Upgrade II phase, planed for the next decade. The limits in the storage capacity of the trigger will bring an inverse relation between the amount of particles selected to be stored per event and the number of events that can be recorded, and the background levels will raise due to the enlarged combinatorics. To tackle both challenges, we propose a novel approach, never attempted before in a hadronic collider: a Deep-learning based Full Event Interpretation (DFEI), to perform the simultaneous identification, isolation and hierarchical reconstruction of all the heavy-hadron decay chains per event. This approach radically contrasts with the standard selection procedure used in LHCb to identify heavy-hadron decays, that looks individually at sub-sets of particles compatible with being products of specific decay types, disregarding the contextual information from the rest of the event. We present the first prototype for the DFEI algorithm, that leverages the power of Graph Neural Networks (GNN). This paper describes the design and development of the algorithm, and its performance in Upgrade I simulated conditions.}




\maketitle

\clearpage

\section{Introduction}
\label{sec_introduction}

The Large Hadron Collider beauty Experiment (LHCb) is one of the four large experiments 
at the proton-proton collider LHC, at CERN~\cite{LHCb:2008vvz}. 
It is dedicated to the study of beauty (b) and charm (c) hadron decays, 
performing high-precision measurements to test the validity of the Standard Model (SM) of particle physics and identify possible signatures of the presence of physics beyond the SM. To push the precision frontier, LHCb needs to record as many heavy-hadron decays as possible. One way to increase that quantity for a given period of data collection is to increment the average number of proton-proton collisions that happen in each event (bunch crossing). During the LHC Run 1 and Run 2 periods, between 2010 and 2018,
each LHCb event contained an average of around one visible proton-proton collision,
producing a flow of tens of particles to be reconstructed. The experiment has now undergone its Upgrade I, with the installation of new sub-detectors and a new data-collection software to allow the processing of events with around five visible proton-proton collisions each. These will be the conditions for the ongoing Run 3 and for the future Run 4. In a decade from now, the Upgrade II~\cite{LHCb:2018roe} of LHCb will prepare the experiment to face another ten-fold increase in proton-collision multiplicity~\cite{Albrecht:2653011} to fully exploit the High-Luminosity (HL-LHC) Phase of the LHC during Runs 5 and 6. The approximated expected object multiplicities per event in the different conditions are shown in Table~\ref{tab_event_anatomy}. Beyond upgraded sub-detectors, the much larger event complexities bring unprecedented challenges to LHCb, both for data-collection and for the eventual measurements. New strategies need to be devised and implemented to tackle those challenges and hence maximise the future physics reach of the experiment.

\begin{table*}[h]
\begin{center}
\begin{minipage}{\textwidth}
\caption{Approximate average quantities per event for the different LHCb run conditions, as estimated from the simulation used in this work. Only objects in LHCb geometrical acceptance are considered.}\label{tab_event_anatomy}%
\begin{tabular*}{\textwidth}{@{\extracolsep{\fill}}llllll@{\extracolsep{\fill}}}
\toprule
LHCb period & Num. vis. pp collisions & Num. tracks & Num. b hadrons & Num. c hadrons \\
\midrule
Runs 1-2 & $\sim1$ & $\sim50$ & $\ll1$ & $\ll1$ \\
Runs 3-4 (Upgrade I) & $\sim5$ & $\sim150$ & $\ll1$ & $\sim1$ \\
Runs 5 (Upgrade II) & $\sim50$ & $\sim1000$ & $\sim1$ & $\sim5$ \\
\botrule
\end{tabular*}
\end{minipage}
\end{center}
\end{table*}

So far, the entire data flow of the LHCb experiment has been based on an exclusive approach, \textit{i.e.} it was sufficient for a set of particles to be compatible with a certain type of decay to be identified as a signal candidate.
While this approach has its merits, it ignores in its selection process all the remaining particles produced in the collision, which contain important information on the underlying physics process.
Exceptions to this exclusive approach are found in flavour tagging algorithms~\cite{Fazzini:2018dyq} and isolation studies~\cite{LHCb:2017rmj,LHCb:2017rln}.
However, both cases look at the rest of the event in relation to a specific candidate, \textit{e.g.} flavour tagging aims at inferring the flavour of the heavy-hadron associated to a given signal candidate.
While technically very challenging, significantly more information could be gained by an inclusive study of all the particles in the event. 
This would not only add discriminating power to disentangle true signal decays from multiple sources of background, but would also allow to identify and separate groups of particles corresponding to multiple heavy-hadron decays in the event, all of which can be used for subsequent physics analyses.
The gains of such an inclusive approach compared to the individual study of signal candidates become stronger with increasing event complexities, as the larger combinatorics problem makes it more complicated to identify and isolate signals.


The individual study of heavy-hadron decays is also at the core of the LHCb strategy for data collection. The trigger of the experiment aims at discerning between events that contain a signal decay and those that don't, by means of a combination of exclusive and partially inclusive~\cite{BBDTTopo,LHCb-PROC-2015-018} particle selections. In past LHC runs, the disk space available to store the information for the selected events was large enough to allow persisting all the objects in the event in many cases. This gave the flexibility to study offline other particles than the ones that compose the signal candidate that triggered the event, which is a crucial feature for signal-background separation in many analyses and to allow the study of modes not considered when the trigger selections were made. This situation is completely different in the HL-LHC era. First of all, the fraction of events containing decays of interest will saturate to around 100\%, with each event typically containing several heavy-hadron decays. Second, the event sizes will be much larger than in the past due to the increased particle multiplicity. This implies that the potential datasets to be collected are huge, while the available disk space is limited and imposes tight constraints. A trigger strategy based on selecting events in those conditions necessarily leads to a signal inefficiency, impacting the potential physics reach of the experiment.
Consequently, the trigger paradigm needs to shift from deciding 
``which events are interesting?'' to ``which parts of the event are interesting?''. 
Minimising the average event size, will directly translate into maximising the number of events LHCb can record. When doing so, the trigger needs to ensure that the relevant particles (those produced in heavy-hadron decays) are amongst those to be kept for offline analysis, otherwise also impacting the potential physics reach. These problems are already partially present in the current LHCb Upgrade I, as anticipated in ref.~\cite{Fitzpatrick:1670985}. In preparation, LHCb has developed a framework that allows the persistency of part of the event information at the technical level~\cite{Aaij:2019uij} (for example, the persistency of the set of reconstructed particles associated to the same proton-proton collision point as a signal candidate). However, at present there is no nominal strategy in LHCb to systematically select which parts of the event may be interesting for physics analysis. This is a very complicated task affected by large particle combinatorics and a huge variability of types of signal decays.

To tackle the previous challenges, we propose a new algorithm to perform a Deep-learning based Full Event Interpretation (DFEI) at LHCb.
This innovative approach, which targets an inclusive analysis of the entire event, represents a shift of paradigm with important applications both at the trigger level and at the offline analysis level.
The algorithm takes as input all the reconstructed particles in an event and aims at identifying which of them originate from the decays of heavy-hadrons and at reconstructing the hierarchical decay chains through which they were produced. The possibility to accomplish this difficult task leverages on some of the most recent developments in the field of machine learning. 
At the trigger level, DFEI can identify the part of each event which is interesting for physics analyses, 
allowing to safely discard the rest of the event and hence minimise the storage required. 
As an additional benefit, an automatised identification and classification of the decay chains could eventually replace the need for cut-based exclusive selections that need to be designed and carefully tuned independently for each signal decay type.
At the offline analysis level, DFEI can offer a common tool for physicists to identify and classify the different types of backgrounds contributing to a broad spectrum of possible decays of interest. Leveraging the information from all the correlations in the event can enhance the background rejection power in many cases, increasing the precision of future LHCb measurements.


This document describes the conceptualisation, construction, training and performance of the first prototype of the DFEI algorithm.
The prototype is specialised for reconstructed charged particles produced in beauty-hadron decays. Extensions to include reconstructed neutral particles and charm-hadron decays can be considered in the future. All the studies are done using simulated datasets 
that emulates proton-proton collisions in the LHCb Run~3 environment. These datasets have been produced with a custom simulation framework, and made publicly available to allow future benchmarking. The algorithm is based on a composition of Graph Neural Network (GNN) models, designed to handle the complexity of high-multiplicity events in a computationally-efficient way. Regarding the paper organisation, the state of the art is first presented in Sec.~\ref{sec_relatedwork}. The development of the DFEI prototype is described in Sec.~\ref{sec_methods}, starting with an introduction to GNN models in Sec.~\ref{sec_gnns}, followed by the description of the employed dataset in Sec.~\ref{sec_dataset} for which additional details are provided in App.~\ref{sec_simulation}, the structure of the algorithm in Sec.~\ref{sec_algorithm}, and finally the training in Sec.~\ref{sec_training}. The performance of the algorithm is described in detail in Sec.~\ref{sec_results}. In particular, the quality of the reconstruction is first evaluated at the event level, in Sec~\ref{sec_physics_performance_event_level}, and then at the exclusive-signal level, in Sec~\ref{sec_physics_performance_signal_level}. A timing study is presented in Sec.~\ref{sec_timing} (with additional details provided in App.~\ref{sec_detailed_timing_constraints}). The results are discussed in Sec.~\ref{sec_discussion} and future prospects are presented in Sec.~\ref{sec_future}. Finally, the conclusions are summarised in Sec.~\ref{sec_conclusion}.
\section{Related work}
\label{sec_relatedwork}

Even though the problem addressed in this paper is unique, it shares similarities with a variety of past efforts at the technical and/or scientific level.
In this section, we do a review of those efforts and put our approach in context in the field.

The first and so far only use of a machine learning based approach on the full set of reconstructed tracks within LHCb is Ref.~\cite{Likhomanenko:2016tgu}, where the authors employed a probabilistic model based on decision trees for the inclusive flavour tagging of signal beauty hadrons. The combined processing of all the event information demonstrated better results compared to a combination of more classical flavour tagging algorithms, each using only a subset of the reconstructed particles in the event. The task of flavour tagging is much simpler than the explicit decay chain reconstruction attempted by DFEI. Regarding isolation tools, past LHCb efforts~\cite{LHCb:2017rmj,LHCb:2017rln} are restricted to multivariate classifiers that aim to predict whether individual particles from the rest of the event originate or not from the same heavy-hadron decay as a signal candidate. The decision is based on a combination of features from the signal candidate and the extra particle, fully disregarding any correlation with the other particles in the event. Concerning trigger-oriented applications, the authors in Ref.~\cite{Bourgeois:2018nvk} presented a study of the full information in the event in terms of the activity in the different LHCb sub-detectors, hence at a level prior to the reconstruction of the stable particles, which are considered as input in DFEI. Using machine learning techniques, they successfully managed to predict the number of reconstructible proton-proton collisions per event\footnote{The DFEI algorithm assumes the proton-proton collision points have already been reconstructed, and usses information of their measured positions as input, as discussed in the following sections.}. They also studied the possible classification of events between those containing (at least) one b-hadron decay and those that don’t, but this turned out to be a very complicated task when looking only at the sub-detector activity information.


Regarding other LHC experiments, a type of full event reconstruction is done in CMS~\cite{CMS:2017yfk} and ATLAS~\cite{ATLAS:2017ghe}, through the usage of the particle flow algorithm. The implementation uses all the final state particles for a global event description, significantly improving the performance of 
jet reconstruction with respect to the previous baseline that used basic geometric cones to cluster particles. In order to further improve the performance, an approach with a GNN~\cite{Pata:2021oez} was proposed in CMS that takes as an input all particles of an event and predicts variables such as particle identification and transverse momentum of each particle.
While similar to DFEI at a technical level, the particle flow algorithm does not attempt to reconstruct explicitly the decay chains for all the relevant decays of interest.

The task of decay-chain reconstruction is conceptually close to the hierarchical reconstruction of jets, for which a variety of algorithms based on GNN were developed~\cite{ATLAS:2022rkn,Huang:2023ssr,Ma:2022bvt,Mokhtar:2022pwm,Ju:2020tbo,Atkinson:2022uzb,Murnane:2022pmd,Gong:2022lye,Konar:2021zdg,Verma:2021ceh,Dreyer:2020brq,Guo:2020vvt,Qu:2019gqs,Moreno:2019bmu,Shlomi:2020ufi}.
The ultimate goal of those algorithms, however, is typically focused on inferring quantities of the jet overall, for example doing a flavour tagging of the jet to determine the initial particle, and reconstructing the jet to infer its kinematics. The
jet substructure is only studied to the extent in which it's useful for those purposes. The limitations of those algorithms for the task of reconstructing all the ancestors in particle decay chains are reviewed in detail in Ref.~\cite{Kahn_2022}.


The effort which is closest to the one presented in this paper is done at the Belle II experiment, where the FEI algorithm~\cite{Keck:2018lcd} was developed for exclusive tagging of B-decays. This constitutes a similar approach as the one presented in this paper but with a different goal and in a simpler environment. As Belle II is a hermetic detector situated at an 
electron-positron collider, the event is a fully reconstructible system with known initial states and significantly less tracks, making the task of inference less challenging. In addition, only two species of b hadrons are studied, $B^0$ and $B^+$ mesons\footnote{Charge conjugation is implied throughout this paper.}, while LHCb is interested in all b-hadron species 
(for example $B_s$ and $B_c$ mesons, and $\Lambda_b$ baryons) as well as c-hadron decays. From a probabilistic point of view, the FEI algorithm at Belle II is based on a fixed set of different boosted decision tree classifiers, one for each considered decay type.
This approach would be unfeasible at LHCb, given the much larger variability in terms of different signal decay topologies, further augmented by the fact that a fraction of the particles produced in the decays may fall outside the LHCb geometrical acceptance, and hence not be reconstructed in the detector. Recently, an extension to the FEI algorithm based on GNN was proposed~\cite{Tsaklidis:2122,Kahn_2022}, showing a better performance than the previous implementation. This resembles the approach presented in this paper, but in a very different environment, as has been discussed.

As exemplified by the previous efforts, GNNs have become popular to replace other machine learning algorithms within particle physics experiments ~\cite{Ju:2020xty, Shlomi_2021}, as they can naturally capture the structure and spatial sparsity of the problem. A challenge however is the GNN's performance in deployment, such as in real-time computing for trigger purposes. Achieving a fast inference with GNNs would require sparse operations and standards of representing such operations in protocols that would allow the automatic optimisation of the networks. This is a matter of broad interest and front-line research. Very recently, there have been multiple successful efforts in this direction within other CERN experiments ~\cite{Thais:2022iok,Que:2022kmo,Elabd:2021lgo,Heintz:2020soy,Iiyama:2020wap,Ju:2021ayy,Pata:2022wam}, for example by reducing the complexity of the networks and using FPGAs or GPUs as hardware accelerators.

%
%
%
%
%
%
%
\section{Methods}
\label{sec_methods}

\subsection{Usage of Graph Neural Networks}
\label{sec_gnns}

Machine learning and especially neural networks usage in particle physics has been growing exponentially in the last decade~\cite{Albertsson:2018maf}.
The major motivation to explore new and increasingly complex
machine learning techniques is to optimally incorporate the structure of the underlying problem into the model itself. This includes incorporating variable input sizes, representing different types of connections between inputs and embedding invariances into the architecture.
Graph Neural Networks are a class of neural networks built around the concept of a graph, which is an unordered and variable-sized collection of nodes ($v\in V$),
edges connecting those nodes ($e\in E$), and possibly a vector of graph-level features ($\mathbf{u}$).
The relations between the nodes occur in a
high-dimensional, non-trivial latent space, allowing for a more complete description of the data. This architecture is especially well fitted to capture problems with sparse connections and invariance under input permutation, as is the case for the set of reconstructed particles in a collision event.

\begin{figure}
    \centering
    \includegraphics[width=\linewidth]{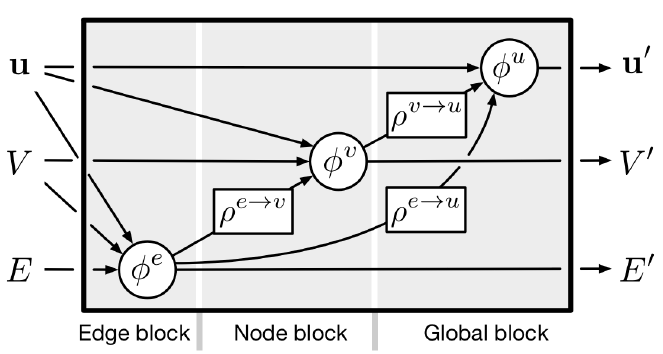}
    \caption{Graph processing block at each message-passing step, as presented in Ref.~\cite{47094}.}
    \label{fig_GN_block}
\end{figure}

In general, GNNs implement ``graph-to-graph'' transformations, by the application of multiple layers that operate on the graph constituents. At each layer, input vectors of features at the node, edge and/or graph level are used and returned to the next layer, with the output of the last layer fulfilling the goal of a certain task. This work is based on the usage of message-passing GNNs, in which the information is propagated through the graph at each layer by exchanging information between adjacent nodes. Specifically, we use the so-called full GN block in Ref.~\cite{47094}, depicted in Fig.~\ref{fig_GN_block}. This block is composed of three feature-update functions, $\phi^v$, $\phi^e$ and $\phi^u$, and three information-aggregation functions, $\rho^{e\to v}$, $\rho^{e\to u}$ and $\rho^{v\to u}$. Each of the three update functions is implemented by a multilayer perceptron (MLP), and the aggregation functions are piece-wise summations.


These blocks are then applied multiple times, and what is returned is a new representation of the graph, the parts of which - nodes, edges or the entire graph - are then up to interpretation or classification as further described in Sec.\ref{sec_algorithm}.



\subsection{Dataset}
\label{sec_dataset}

The DFEI prototype is trained on simulated data. Since the LHCb simulation samples are restricted to internal member access only, and there is no publicly available dataset that fully captures the essence of the problem at hand, we have created a new simulation environment and produced datasets which we have made publicly available~\cite{dfei_dataset}. The datasets are generated with PYTHIA8~\cite{Bierlich:2022pfr} and EvtGen~\cite{Ryd:2005zz}, replicating the particle-collision conditions expected for the LHCb Run 3.
In addition, an approximate emulation of the LHCb detection and reconstruction effects is applied, as described in App.~\ref{sec_simulation}. 
In the generated dataset, each event is required to contain at least one b-hadron, which is subsequently allowed to decay freely through any of the standard decay modes present in PYTHIA8. In these conditions, around 40\% of the events contain more than one b-hadron decay within LHCb acceptance, and the maximum observed b-hadron decay multiplicity is five.
All the studies presented in this paper refer only to reconstructed particles that have been produced inside the LHCb geometrical acceptance and in the Vertex Locator region (as defined in App.~\ref{sec_simulation}). Other particles are not considered, which also implies they are not included as part of the ground truth heavy-hadron decay chains.

A total of $100\,000$ simulated events have been used to develop this first prototype of the DFEI algorithm. They are divided in: training dataset ($40\,000$ events), test dataset ($10\,000$ events) and evaluation dataset ($50\,000$ events).
In addition to this inclusive dataset, several other smaller samples (of few thousand events each) have also been generated simulating specific signal decay types.
These decay types have been chosen to be representative of the most common signal topologies studied in physics analyses at LHCb, and are used to evaluate the performance of DFEI focused on typical use cases. These samples contain only events in which all the particles originating from each of the considered exclusive decays have been produced inside the LHCb geometrical acceptance and in the Vertex Locator region. 


The input features used in the DFEI GNN modules 
are described in the following. Regarding geometrical variables, a cartesian right-handed coordinate system is adopted, with the $z$ axis along the beam, the $y$ pointing upwards and the $x$ axis parallel to the horizontal.
\begin{itemize}
\item Node variables:
\begin{itemize}
    \item Transverse momentum ($p_T$): component of the three-momentum transverse to the beamline.
    \item Impact parameter (IP) with respect to the associated primary vertex (PV): distance of closest approach between the particle trajectory and its associated primary vertex (proton-proton collision point), defined as the one with the smallest IP for the given particle amongst all the primary vertices in the event.
    \item Pseudorapidity ($\eta$): spatial coordinate describing the angle of a particle relative to the beam axis, computed as $\eta=\mathrm{arctanh}(p_z/\|\vec{p}\|)$.
    \item Charge ($q$): since only charged reconstructed particles are considered, the charge can only take the values 1 or -1.
    \item $O_x$, $O_y$, $O_z$: cartesian coordinates of the origin point of the particle.
    \item $p_x$, $p_y$, $p_z$: cartesian coordinates of the three-momentum of the particle.
    \item $PV_x$, $PV_y$, $PV_z$: cartesian coordinates of the position of the associated primary vertex.
\end{itemize}
\item Edge variables:
\begin{itemize}
    \item Opening angle ($\theta$): angle between the three-momentum directions of the two particles.
    \item Momentum-transverse distance ($d_{\perp\vec{P}}$): distance between the origin point of the two particles projected onto a plane which is transverse to the combined three momentum of the two particles.
    \item Distance along the beam axis ($\Delta_z$): difference between the $z$-coordinate of the origin points of the two particles.
    \item $FromSamePV$: boolean variable indicating whether the two particles share the same associated primary vertex.
    \item $IsSelfLoop$: boolean variable indicating whether the edge is connecting a particle with itself or not (i.e. it connects two different particles).
\end{itemize}
\end{itemize}
\subsection{Structure of the algorithm}
\label{sec_algorithm}

To simplify the problem and improve the scalability of the algorithm, a sequential approach is adopted: several event pre-filtering steps are applied before the decay chain reconstruction is performed. Each collision event is transformed into a graph, where the charged reconstructed particles are represented as nodes and the relations between them are represented as edges. Edges are established between particles that either share the same associated primary vertex or have an opening angle smaller than a given threshold.
Requiring that the edge selection keeps 99\% of the connections between particles that originate from the same b-hadron decay corresponds to choosing a threshold value for $\theta$ of 0.26 rad. This requirement removes around 11\% of all the other connections.
A further tuning of this parameter goes beyond the scope of this paper, which privileges a loose preselection in order not to compromise the subsequent performance of the algorithm.

\begin{figure}[h]%
\centering
\includegraphics[width=174pt]{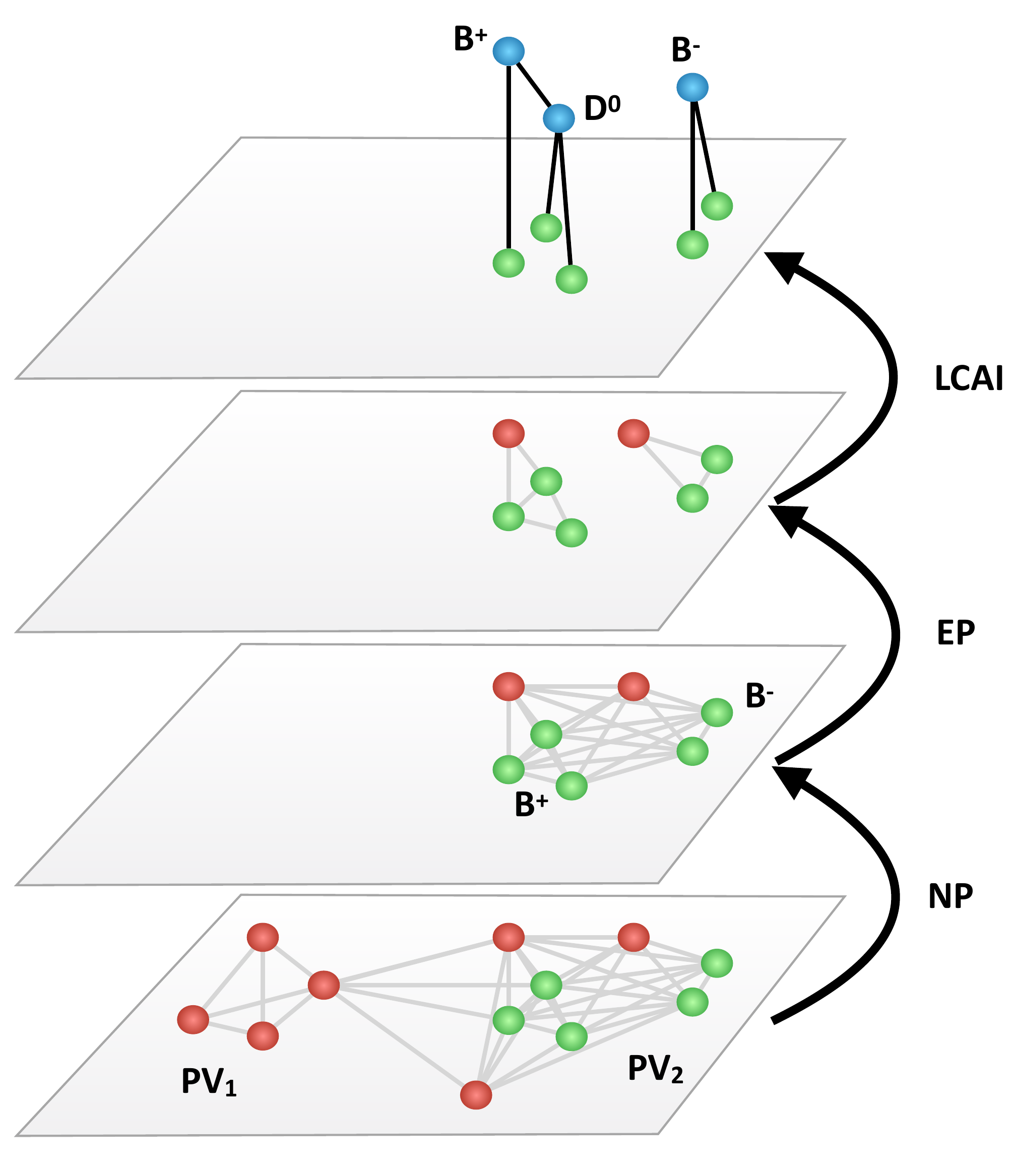}
\caption{Schematic representation of an event processing by the algorithm. Green (red) graph nodes represent particles originated in the decay chain of a b-hadron (from the rest of the event). The reconstructed ancestors are represented in blue.}\label{fig_algorithm_diagram}
\end{figure}

The input graph is passed subsequently through three GNN modules, built using the \texttt{graph\_nets} library~\cite{47094}. The modules are schematically represented on Fig.~\ref{fig_algorithm_diagram} and described in the following. The input features used by each module are specified in Table~\ref{tab_algorithm_input_variables}.

\begin{table*}[h]
\begin{center}
\begin{minipage}{\textwidth}
\caption{Input variables used by each of the DFEI modules. In addition, the total number of stable particles per event ($NumParts$) is included as an input variable. Where relevant, edges connecting two different particles are differentiated from edges connecting a particle with itself (self-loops) by the boolean variable $IsSelfLoop$.}
\label{tab_algorithm_input_variables}
\begin{tabular*}{\textwidth}{@{\extracolsep{\fill}}llll@{\extracolsep{\fill}}}
\toprule
Model & Node variables & Edge variables & Global variables \\
\midrule
NP    & $p_T$, $IP$, $\eta$, $q$ & $FromSamePV$, $\theta$, $d_{\perp\vec{P}}$, $\Delta_z$, $IsSelfLoop$ & $NumParts$    \\
EP    & $p_T$, $IP$, $\eta$, $q$ & $FromSamePV$, $\theta$, $d_{\perp\vec{P}}$, $\Delta_z$ & $NumParts$    \\
LCAI    & $O_{(x,y,z)}$, $p_{(x,y,z)}$, $PV_{(x,y,z)}$, $q$ & $FromSamePV$, $\theta$, $d_{\perp\vec{P}}$, $\Delta_z$ & $NumParts$    \\
\botrule
\end{tabular*}
\end{minipage}
\end{center}
\end{table*}

\begin{itemize}
    \item \textbf{Node pruning (NP)}. The first GNN module has the goal of removing most of the particles (nodes) that that have not been produced in the decay of any b-hadron. It mostly exploits the fact that particles produced in the decay of a b-hadron typically have large IP and $p_T$ values. Since the main contributing factor to the prediction of each node in this case comes from the same node's features, self-loop connections are included in the graphs. The model is trained using binary cross-entropy to predict whether a node originates from beauty hadrons or not. Nodes with an output score below a certain threshold are removed from the graph.
    \item \textbf{Edge pruning (EP)}. The graph in output of the previous step still has a large number of edges, which are further reduced by a second GNN module. This one has the aim of removing edges between particles that don’t share the same beauty-hadron ancestor. Amongst other relations, this exploits the fact that particles coming from the same b-hadron decay tend to be closer in space and their three-momenta tend to form a small opening angle.
    The model is trained using binary cross-entropy to predict whether an edge connects two particles that originate from the same beauty hadron decay or not. Edges with an output score below a certain threshold are removed from the graph.
    \item \textbf{Lowest common ancestor inference (LCAI)}. Finally, a third GNN module takes the output of the previous algorithm, and aims at inferring the so-called “lowest common ancestor” of each pair of particles (a technique similar to the recently proposed LCA-matrix reconstruction for the Belle II experiment~\cite{Tsaklidis:2122}).
    The limited coverage of the LHCb geometrical acceptance and the fact that only charged reconstructed particles are considered in this prototype implies that a large fraction of the decay chains can only be partially reconstructible. To circumvent this limitation, the target decay chains for this prototype are not the ones output by the PYTHIA8 simulation but a ``topological'' version of them, constructed from the separable decay vertices in the decay chain. In practice, this amounts to a transformation of the ground truth decay chain, removing the ancestors that either correspond to very-short-lived resonances or don’t have enough charged-particle descendants to allow the formation of a vertex. From a technical perspective, the GNN module performs a multi-class classification on the edges, outputting a score associated to the ``topological'' LCA relation between the two connected particles, \textit{e.g} particles that share the same mother will have 1st order LCA (class-1), particles that have the same grand-mother will have 2nd order LCA (class-2), etc..
    The fraction of edges with a ground-truth order larger than 3 in the simulation is very small, so the target classes considered are class-1, class-2 and class-3. In addition, a class with an LCA value of 0 is included (class-0), to identify the case in which the two particles don't originate from the same decay chain. As a side product of the addition of this last class, the LCAI provides a final step of node filtering, by allowing to remove fully-disconnected particles (those whose edges are all predicted to have an LCA value of 0).
\end{itemize}

Each of the previous modules uses independent MLPs for the node-, edge- and global-update functions introduced in Sec.~\ref{sec_gnns}. Each MLP is composed of a certain number of layers, all of which have the same latent size. The number of GN block iterations is also configured separately for each module. The hyperparameters chosen for this prototype are written in Table~\ref{tab_model_and_training_hyperparameters}.

The output of the DFEI processing chain can be directly translated into a set of selected charged reconstructed particles and their inferred ancestors, with the predicted hierarchical relations amongst them.


\subsection{Training}
\label{sec_training}

The training is done in stages, following the algorithm sequence. Each model is trained in a supervised way, using a weighted softmax cross entropy as loss function, where the weights (corresponding to the inverse of the number of elements in each true class) compensate for the imbalance across classes present in the dataset.
The minimisation is done using Adam with the hyperparameter configuration reported in Table~\ref{tab_model_and_training_hyperparameters}.

\begin{table*}[h]
\begin{center}
\begin{minipage}{\textwidth}
\caption{Hyperparameters used in the construction and training of the different GNN modules.}
\label{tab_model_and_training_hyperparameters}%
\begin{tabular*}{\textwidth}{@{\extracolsep{\fill}}lllllll@{\extracolsep{\fill}}}
\toprule
Model & \# layers  & Latent size & \# GN blocks & Batch size & Learning rate & \# training steps\\
\midrule
NP    & 3   & 50  & 3 & 32 & $5\cdot10^{-4}$ & 500 \\
EP    & 4   & 100   & 5 & 32 & $10^{-4}$ & 500 \\
LCAI    & 5   & 100  & 5 & 128 & $10^{-3}$ & $2\cdot10^{5}$ \\
\botrule
\end{tabular*}
\end{minipage}
\end{center}
\end{table*}

Thresholds are defined for the output score of the NP and EP models, as those resulting into a $\sim99\%$ efficiency of selecting the desired nodes and edges, respectively. This loose requirement is chosen to minimise the potential negative impact on the performance of the subsequent steps. The working point corresponds to a $\sim70\%$ background rejection power for nodes from the NP algorithm
and a $\sim68\%$ background rejection power for edges from the EP algorithm.
In this setup, the ROC AUC for the NP module is 0.977, and the one for the EP module is 0.974. A consistent performance is observed between the training and test samples, showing no overtraining of these modules. The average reduction of the total event size after each processing step is shown in Tab.~\ref{tab_event_reduction_factors}.

\begin{table}[h]
\begin{center}
\begin{minipage}{174pt}
\caption{Cumulative average efficiencies on the total number of nodes and edges in the graph after each pre-filtering step, illustrating the graph reduction power achieved in each case.}\label{tab_event_reduction_factors}%
\begin{tabular}{@{}lll@{}}
\toprule
Filtering step & Node eff.  & Edge eff. \\
\midrule
Edge pre-selection    & $\sim$100\,\%   & $\sim$89\,\%  \\
NP    & $\sim$29\,\%   & $\sim$6\,\%  \\
EP    & $\sim$27\,\%   & $\sim$2\,\%  \\
\botrule
\end{tabular}
\end{minipage}
\end{center}
\end{table}

The training of the LCAI module requires significantly more training iterations than the previous steps, given the much higher complexity of the task. A certain level of overtraining is found for the least populated classes, and the training is stopped once the average classification accuracy for the test sample reaches a plateau.
Since the goal of this paper is demonstrating the feasibility of the approach by presenting a first working prototype, rather than obtaining the maximum possible performance, we leave the improvements in the training as future work.
\section{Results}
\label{sec_results}

In this section, the performance of the current DFEI prototype is described, both at an event level (relevant for trigger) and at an individual-decay-chain level (relevant for trigger and offline analysis).

\subsection{Event-level performance}
\label{sec_physics_performance_event_level}

Different metrics are defined and evaluated in the following to characterise the performance of DFEI at event level, from multiple perspectives.

\begin{figure}[h]%
\centering
\includegraphics[width=0.48\textwidth]{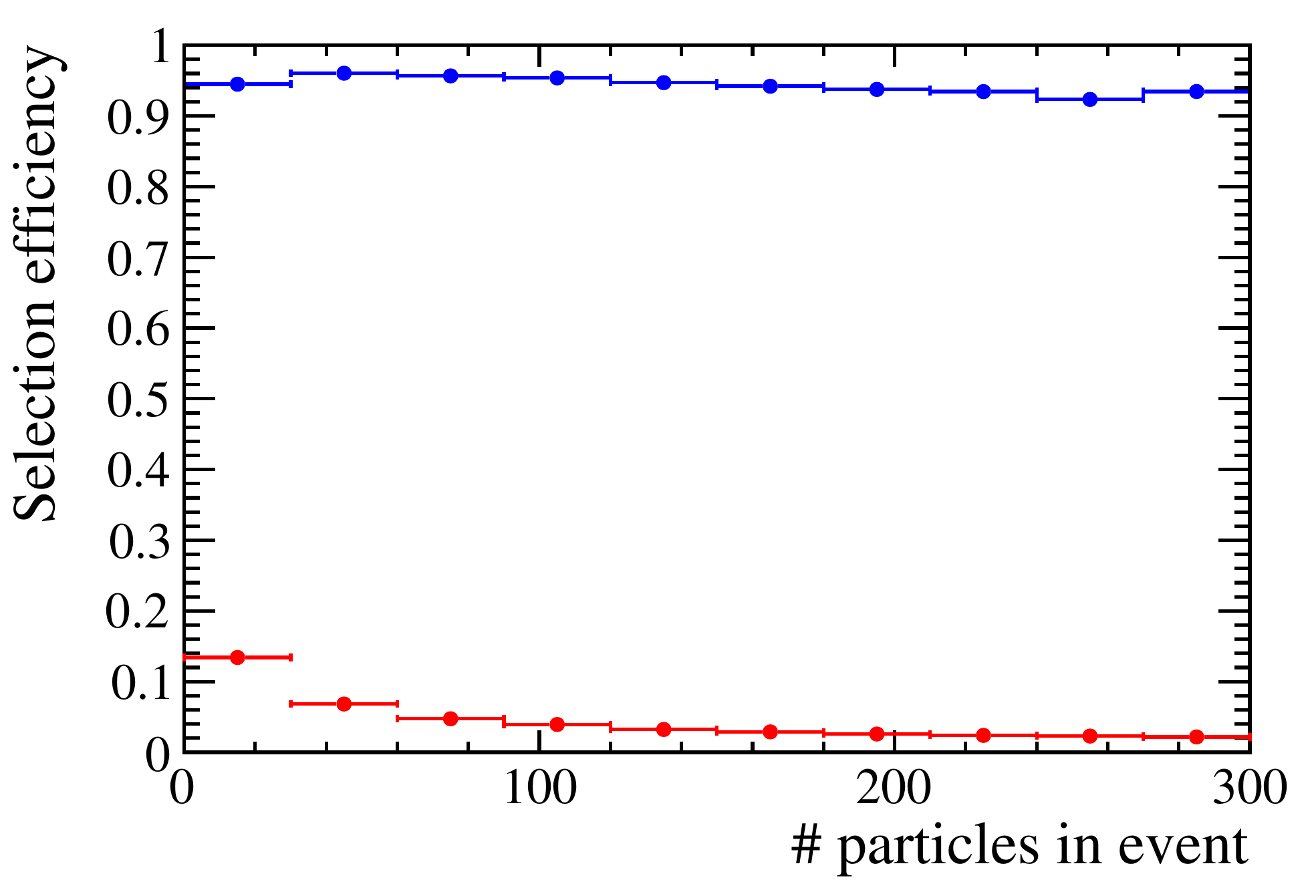}\\
\caption{Average particle-selection efficiency as a function of the total number of particles per event, shown separately for (blue) particles originating from a b-hadron decay and (red) particles from the rest of the event. }
\label{fig_sel_eff_differential_multiplicity}
\end{figure}

\begin{figure}[h]%
\centering
\includegraphics[width=0.48\textwidth]{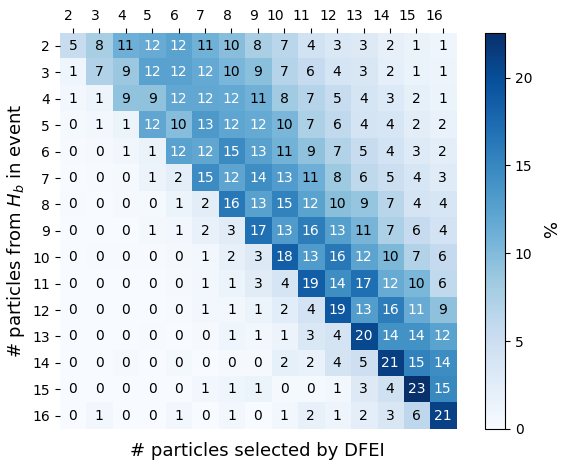}
\caption{Confusion matrix for the true vs. predicted number of particles from b-hadron decays per event, computed in terms of percentages normalised for each row (true value) and shown for the square subregion corresponding to a number of particles between 2 and 16.}
\label{fig_conf_matrix_num_signal_particles}
\end{figure}

\paragraph{Event-size-reduction capabilities.} Three different quantities are studied, as a function of the particle multiplicity per event: efficiency of selecting particles from a b-hadron ($H_b$) decay, efficiency of rejecting particles from the rest of the event (background), and total number of selected particles in the event. The obtained values are shown in Fig.~\ref{fig_sel_eff_differential_multiplicity} and Fig.~\ref{fig_conf_matrix_num_signal_particles}. The average efficiency for selecting particles truly produced in b-hadron decays is 94\%, and the average background rejection power is 96\%. An almost flat response as a function of the total number of particles in the event is found for the selection of particles from b-hadron decays. The background rejection power mildly  increases with the multiplicity. The average number of selected particles per event is $\sim$10, from the initial number of $\sim$140.
A good event reduction is obtained irrespectively of the number of particles originating from b-hadron decays per event, as demonstrated by the linear behavior of the confusion matrix presented in Fig.~\ref{fig_conf_matrix_num_signal_particles}. For the set of selected particles per event, an average purity of 60\% is found, defined as the number of selected particles that truly originate from b-hadron decays over the total number of selected particles.

\paragraph{Quality of the decay-chain reconstruction.}

Apart from helping in background suppression in offline analysis, being able to accurately reconstruct and classify the decay chains in an event can allow DFEI to allow a further level of automation to the LHCb trigger, as introduced in Sec.~\ref{sec_introduction}.

\begin{figure}[h!]%
\centering
\includegraphics[width=0.45\textwidth]{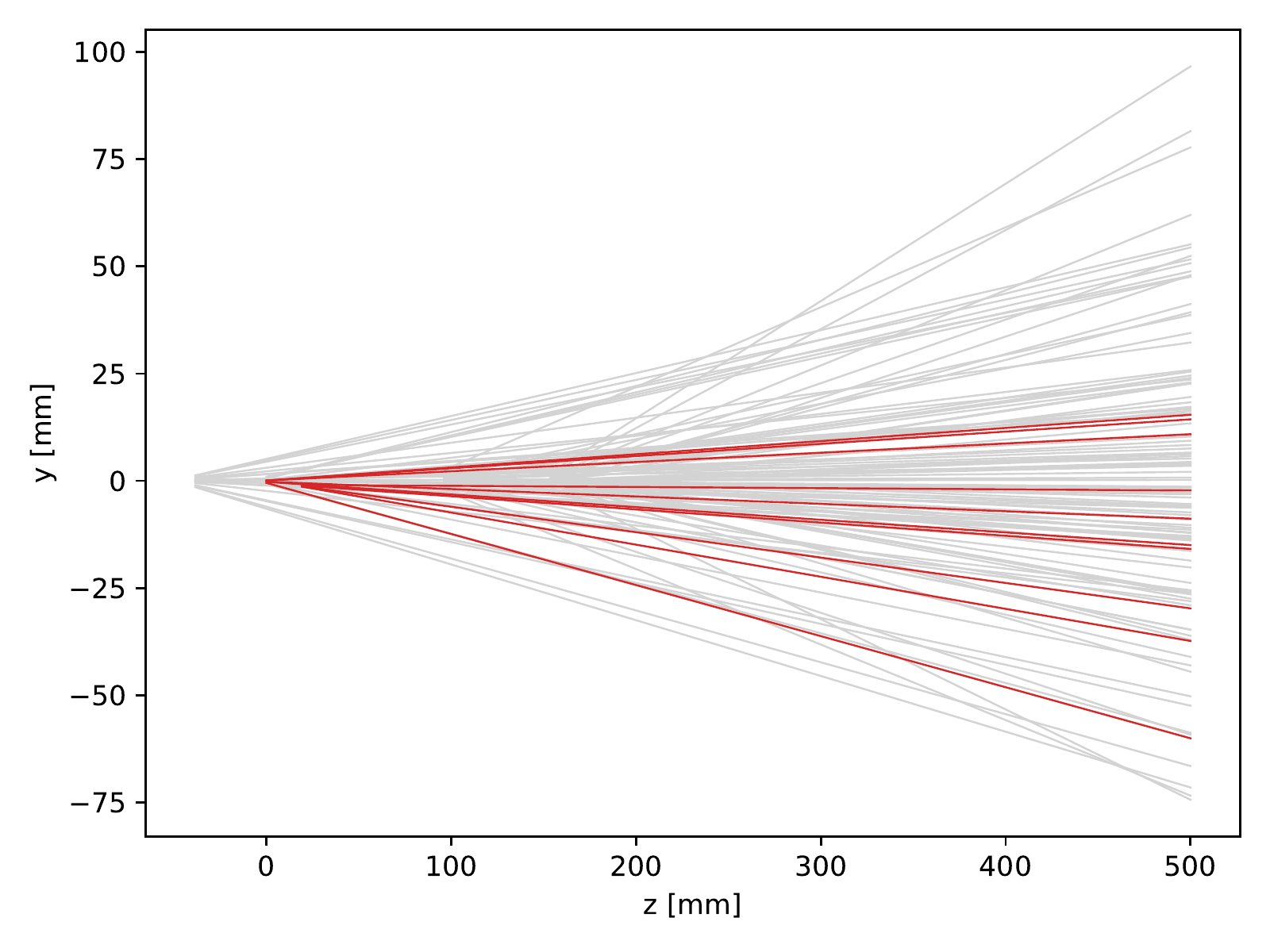}
\caption{Example of a PER from the evaluation dataset. Two-dimensional view of the charged reconstructed particle trajectories in the proton-proton interaction region. Red lines represent particles produced in b-hadron decays, that DFEI has correctly selected, and gray lines represent particles from the rest of the event, that DFEI has correctly removed.}\label{fig_PER_example_detector}
\end{figure}

\begin{figure}[h!]%
\centering
\includegraphics[width=0.48\textwidth]{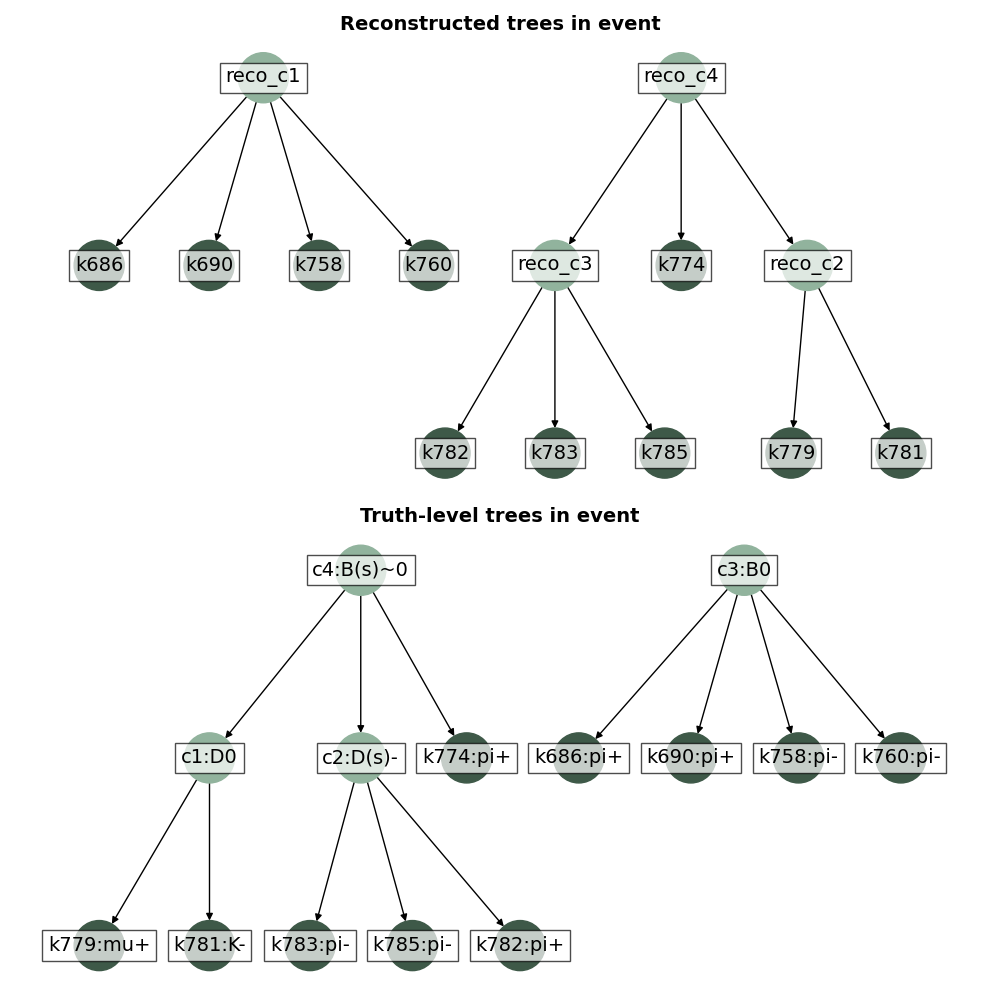}
\caption{Example of a PER from the evaluation dataset, same as in Fig.~\ref{fig_PER_example_detector}. The (top) reconstructed and (bottom) ground truth b-hadron decay chains in the event are shown. Apart from the reconstructed particles produced in those decays, the event contains 106 particles from the rest of the event (not shown for simplicity), all of which are correctly removed by DFEI. The dark-green (light green) circles represent the reconstructed particles (topological ancestors). The key (k) numbers correspond to unique identifiers for each reconstructed particle produced in the simulation. The cluster (c) numbers correspond to unique identifiers assigned to each ancestor during the construction of the decay chains. The true identity of the particles is shown in the ground-truth case.}
\label{fig_PER_example_tree}
\end{figure}

A first metric that can serve for characterising the overall understanding of the event in this regard, and be used for benchmarking purposes, is the fraction of events in which DFEI achieves a perfect event reconstruction (PER). For an event to fulfill this condition, all the b-hadron decays in the event need to have been found, all the charged reconstructed particles produced in them been selected, the associated ``topological'' decay chains been exactly reconstructed, and all the particles from the rest of the event been removed. An example of a PER case found by DFEI in the evaluation dataset is shown on Fig.~\ref{fig_PER_example_detector} and Fig.~\ref{fig_PER_example_tree}, from the points of view of the reconstructed-particle filtering and of the ancestor-chain reconstruction, respectively. The average fraction of PER found in the evaluation dataset is $(2.14\pm0.07)\%$.

It should be noted that the PER is an extremely challenging case, and that even a partially-good reconstruction can be used for trigger purposes. For example, the selection of extra particles from the rest of the event will break the conditions for a PER, but won't impact the efficiency for selecting all the particles produced in b-hadron decays.

\subsection{Decay-level performance}
\label{sec_physics_performance_signal_level}

The performance shown in the previous section refers inclusively to all the heavy-hadron decays per event and, for each of them, it considers an average over all the known b-hadron species and their known decay types. In this section, the DFEI output is processed to obtain predictions referred to individual decays. For this, true decay chains of a certain type are first searched for in the simulation datasets, and the sub-set of reconstructed particles produced in them identified. Then, the DFEI output for the events containing those decays is studied, focusing on the prediction for the identified sub-set of reconstructed particles. The decay-chain reconstruction achieved by DFEI can subsequently be classified according to the following mutually-exclusive categories:
\begin{itemize}
    \item \textbf{Perfectly reconstructed}: all the reconstructed particles originating from the b-hadron decay have been predicted to be part of the same connected sub-graph, which is disconnected from all the other particles in the event, and the ``topological'' ancestor decay chain has been perfectly reconstructed.
    \item \textbf{Wrong hierarchy}: same as before, but there is at least one mistake in the reconstruction of the ``topological'' ancestor decay chain.
    \item \textbf{Not isolated}: all the reconstructed particles originating from the b-hadron decay have been predicted to be part of the same connected sub-graph, but there is at least one extra particle from the rest of the event which is also contained in that sub-graph.
    \item \textbf{Partially reconstructed}: not all the reconstructed particles originating from the b-hadron decay have been predicted to be part of the same connected sub-graph. It should be noted that this type of reconstruction does not necessarily imply an overall inefficiency for selecting the particles from the b-hadron decay, since they can have been selected in multiple sub-graphs.
\end{itemize}

The decay-level performance is first computed in an inclusive way using the evaluation dataset, by measuring individually the response for all the b-hadron decays contained in the simulation and then taking the average of the performances. The numbers are reported in Table.~\ref{tab_signal_based_reconstruction}. Complementary to the inclusive case, the DFEI response is evaluated in a second stage restricted to specific decay types, by using the additional datasets introduced in Sec.~\ref{sec_dataset}.
The resulting numbers are also reported in Table.~\ref{tab_signal_based_reconstruction}. Those modes are representative of the most typical case studies of LHCb, with the inclusive sample also containing decays to many particles and more complicated decay topologies, for which the reconstruction is more challenging.

\begin{table*}[h]
\begin{center}
\begin{minipage}{\textwidth}
\caption{Decay-level performance of DFEI for the inclusive ($H_b$) case and for several exclusive decay types. In the cases in which the fraction is measured to be zero, the frequentist Wilson upper limit~\cite{Wilson:1927} for a 68\% coverage is provided.}
\label{tab_signal_based_reconstruction}%
\begin{tabular*}{\textwidth}{@{\extracolsep{\fill}}lllll@{\extracolsep{\fill}}}
\toprule
Decay mode & Perfect (\%) & Wrong hierarchy (\%) & Not iso. (\%) & Part. reco. (\%) \\
\midrule
Inclusive $H_b$ decay & 4.6 $\pm$ 0.1 & 5.9 $\pm$ 0.1 & 76.0 $\pm$ 0.2 & 13.4 $\pm$ 0.1 \\
\midrule
$B^{0} \to {K}_{0}^{*}[K \pi] \mu^{+} \mu^{-}$ & 35.8 $\pm$ 0.7 & 19.2 $\pm$ 0.6 &44.9 $\pm$ 0.7 & $<$0.02 \\
$B^{0} \to K^{+} \pi^{-}$ & 38.0  $\pm$ 0.7 & $-$ & 54.7 $\pm$ 0.7 & 7.2  $\pm$ 0.4 \\
$B_{s}^{0} \to D_{s}^{-} [K^{-} K^{+} \pi^{-}] \; \pi^{+}$ & 32.8  $\pm$ 0.7 & 7.1 $\pm$ 0.4 & 53.7 $\pm$ 0.8 & 6.4  $\pm$ 0.4 \\
$B^{0} \to D^{-}[K^+\pi^-\pi^-] D^{+}[K^-\pi^+\pi^+]$ & 22.7 $\pm$ 0.6 & 22.4 $\pm$ 0.6 & 54.9 $\pm$ 0.8 & $<$0.02 \\
$B^{+} \to K^{+} K^{-} \pi^{+}$ & 35.7  $\pm$ 0.7 & 10.2 $\pm$ 0.4 & 46.4 $\pm$ 0.7 & 7.7  $\pm$ 0.4 \\
$\Lambda_{b}^{0} \to \Lambda_{c}^{+} [p K^{-} \pi^{+}] \; \pi^{-}$ & 21.7  $\pm$ 1.0 & 8.9 $\pm$ 0.7 & 36.8 $\pm$ 1.2 & 32.6  $\pm$ 1.1 \\
$B_{s}^{0} \to J/\psi[\mu^+\mu^-] \; \phi[K^+K^-] $ & 26.9  $\pm$ 0.6 & 20.5 $\pm$ 0.5 & 52.5 $\pm$ 0.6 & $<$0.02 \\
\botrule
\end{tabular*}
\end{minipage}
\end{center}
\end{table*}

 The performance evaluated on the exclusive modes is significantly better than the inclusive case, with fractions of perfectly reconstructed decays in the range $20-40\%$. The comparative study of the performance on the different exclusive modes helps to understand which cases are easier or harder for DFEI to reconstruct, and in general to analyse the dependencies of the DFEI response. The most complicated cases are found to be $B^{0} \to D^{-}[K^+\pi^-\pi^-] D^{+}[K^-\pi^+\pi^+]$ (with two three-particle vertices very separated in space, given the long lifetime of the $D^+$ meson), and $\Lambda_{b}^{0} \to \Lambda_{c}^{+} [p K^{-} \pi^{+}] \; \pi^{-}$ (with a single $\pi^+$ that needs to be associated to a spatially-separated three-particle vertex. The difference in performance between the second of the previous decays and $B_{s}^{0} \to D_{s}^{-} [K^{-} K^{+} \pi^{-}]$, which has a similar topology, is due to the $\Lambda_{c}^{+}$ flying more on average than the $D_{s}^{-}$, due to a significantly larger Lorentz boost. The fraction of partial reconstruction is below $10\%$ in all the exclusive cases except for the $\Lambda_{b}^{0} \to \Lambda_{c}^{+} [p K^{-} \pi^{+}] \; \pi^{-}$ decay, which translates into an efficiency for selecting all the reconstructed particles produced in those decays above $90\%$.

\subsection{Timing studies}
\label{sec_timing}



Detailed timing studies and an optimisation of the inference speed of the DFEI algorithm are out of the scope of this paper, and are left for future research. However, a first, simplified, timing study of the current prototype is shown in this section. The first motivation for the study is to understand the scalability of the response with the object multiplicity per event. The second goal is to estimate how the current event-processing rate achievable by the algorithm compares with the requirements to run DFEI in the current Run 3 trigger of LHCb. As explained in App.~\ref{sec_detailed_timing_constraints}, this would imply a processing rate in the ballpark of 500 Hz per computing node (the precise target number would depend on internal LHCb considerations).

\begin{figure}[h]%
\centering
\includegraphics[width=0.48\textwidth]{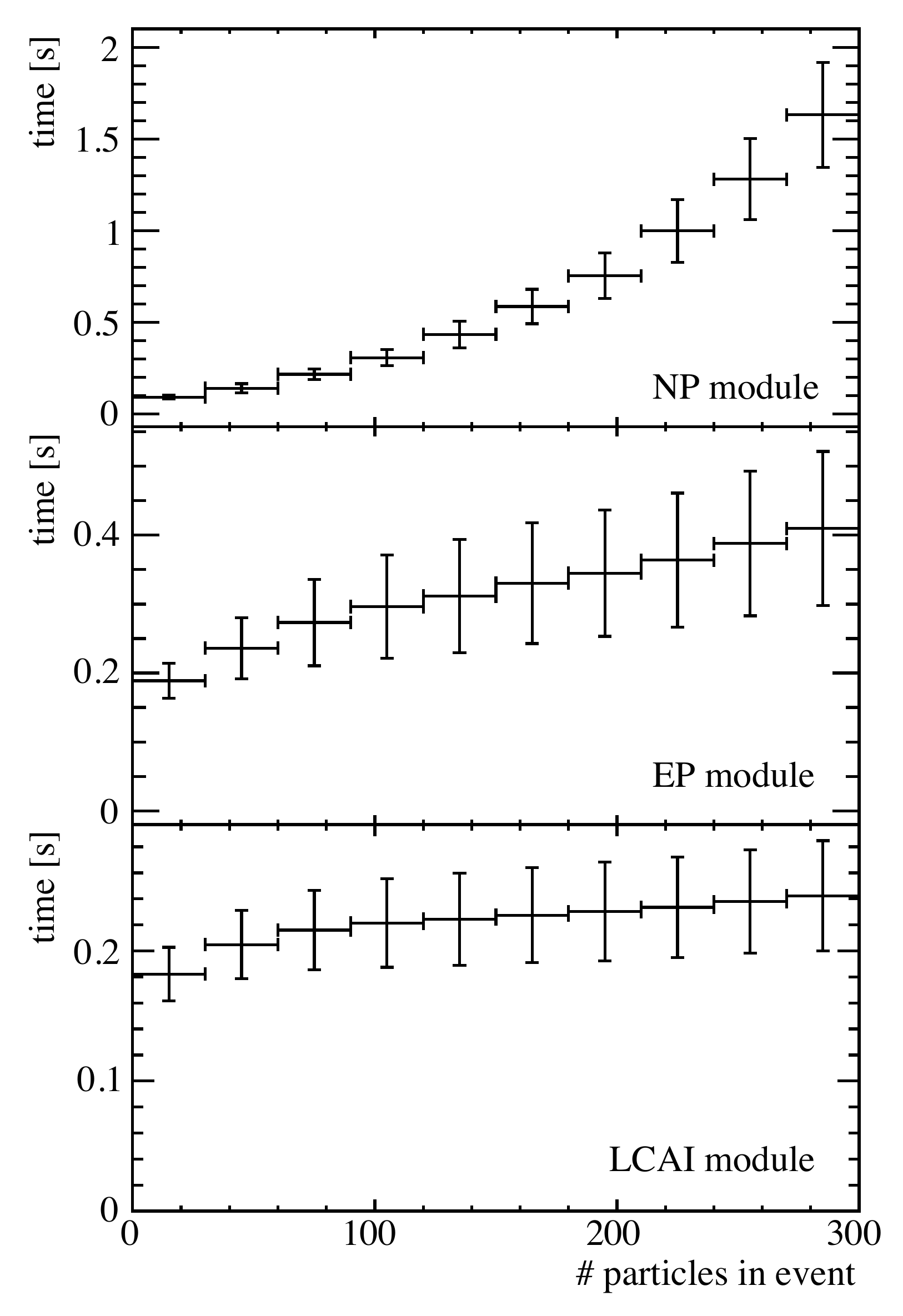}
\caption{Average evaluation time per event of the different DFEI modules as a function of the total number of reconstructed particles in the event. The error bars correspond to the standard deviation.}\label{fig_timing_performance}
\end{figure}

The timing study is done on a CentOS Linux 7 (Core) x86 architecture, using a 2.2 GHz Intel Core Processor (Broadwell, IBRS). No parallelisation scheme is employed. The average computing time required for the evaluation of the NP, EP and LCAI modules as a function of the total number of particles in the event is computed and reported in Fig.~\ref{fig_timing_performance}. In this configuration, the NP is both the slowest module and the one that presents the strongest scaling as a function of the event size, hence the one that can profit the most from a future optimisation in terms of timing. The average of the combined NP + EP + LCAI times is approximately 1 s per event. 

The time needed to create the input graph\footnote{The values of the input features are assumed to be already available at the time DFEI is evaluated, as is the case in the datasets used in this paper.} for each of the three modules and to post-process their output (i.e. filtering nodes and edges and interpreting the predicted LCA values in terms of reconstructed decay chains) is not included in the previous study. From all these auxiliary tasks, the only one that doesn't have a processing time significantly under 1 s is the graph construction of the NP module, that requires an average of 2 s per event.

Taking into account these first studies, a strategy to speed up the full algorithm in order to meet the trigger constraints is outlined in Sec.~\ref{sec_future}.
\section{Discussion}
\label{sec_discussion}

The proposed approach for a multi-heavy-hadron-decay reconstruction of b-hadron decays in a hadronic environment is the first of its kind. To allow the benchmarking of future efforts in this new scenario, all the datasets used for the training and evaluation performance of DFEI have been made publicly available~\cite{dfei_dataset}. In this section, the performance obtained with this first prototype is discussed, in reference to the global context.


On a first stage, the reconstructed-particle selection capabilities can be compared with previous studies in LHCb.
The closest case study, reported in Ref.~\cite{Aaij:2019uij}, considers the sub-set of reconstructed particles that have been selected by a standard LHCb inclusive trigger algorithm, and attempts to discern whether each of the other particles in the event has been produced in the same b-hadron decay or not. By combining vertex-quality requirements and the output of a multivariate algorithm trained on individual-particle features, the authors estimate an approximate selection efficiency for particles from the same b-hadron decay of 90\% for an approximate background rejection power of 90\%. That study is based on official LHCb simulation, which contains material-interaction backgrounds and fake-track backgrounds, not included in the simulated dataset used in this paper. Both simulations, however, aim at representing inclusive b-hadron decays in LHCb Run~3-like conditions. The performance of DFEI (94\% selection efficiency for particles from b-hadron decays and 96\% background rejection power) is similar and numerically higher, within the caveats of the comparison. Most importantly, it shows a powerful discrimination consistently for all the b-hadron decays present in the event at the same time, instead of focusing on an individual decay. It should be noted that the strategy presented in Ref.~\cite{Aaij:2019uij} is not used in production by LHCb. The difference between the two approaches will only increase in the much harsher object-multiplicity conditions expected for LHCb Upgrade II. The almost flat response found in DFEI for the particle-selection efficiencies as a function of the number of particles in the event also suggests good prospects for the Upgrade II conditions.

On a second level, regarding decay-chain reconstruction, DFEI has demonstrated for the first time that this kind of reconstruction can be done successfully both in a hadronic environment and in a multi-decay-chain scenario.
Given the novelty of the approach, the performance at this level can only be partially compared with the one achieved by the FEI algorithm at the Belle~II experiment, and with significant caveats. On one side, as explained in Sec.~\ref{sec_introduction}, the reconstruction in LHCb is a much more difficult task than in Belle~II.
On the other side, the DFEI prototype for LHCb makes use of several, previously introduced, simplifications: omitting particles produced outside the geometrical acceptance, not including neutral reconstructed particles and reconstructing only the ``topological'' decay chains, not the full ones. Keeping the previous caveats in mind, the fraction of perfect decay reconstruction obtained in this paper can be approximately compared to the so-called tag-side reconstruction efficiency determined in Ref.~\cite{Keck:2018lcd} using a Belle simulated dataset, which is of the order of few per cent for semileptonic decays and of few per mille for hadronic decays. The conclusion that can be drawn from this comparison is that DFEI manages a level of reconstruction of decay chains in a hadronic environment which is in the ballpark of that achieved in Belle~(II), hence demonstrating not only the feasibility but also the competitiveness of a Full Event Interpretation approach at the LHC.



Concerning offline analysis applications, a study of different possible types of DFEI reconstruction for specific ground truth decay chains is reported in Sec.~\ref{sec_physics_performance_signal_level}. A technically similar but conceptually different study could be done on a collision dataset, focusing this time on the DFEI prediction for the reconstructed particles output by any standard LHCb analysis preselection. Those preselections aim at identifying particles that are compatible with having been produced in a specific type of decay chain, which are denoted as signal candidates. The DFEI output can be used to classify each signal candidate in one of the following categories: ``signal'' (if the reconstruction matches the expected decay chain), background with a different resonance structure (if the selected reconstructed particles are deemed to be correct but the predicted hierarchy is not), background from decays with extra particles (some of which are not part of the signal candidate) and combinatorial background (where the candidate particles are predicted to originate from multiple sources). This implies that DFEI could virtually be used in every LHCb analysis to suppress/study the different possible types of contributing backgrounds with a potentially-higher background separation power, by leveraging all the information in the event.

\subsection{Future work}
\label{sec_future}

The work in this paper opens the door for multiple future research lines. Natural follow up steps are detailed performance studies on official LHCb simulation and Run 3 collision data. This will allow to assess the impact of the DFEI reconstruction in a broad spectrum of decay distributions, to understand the potential needs for further optimisations/calibrations of the algorithm. Another natural continuation is the extension of the developments and studies to Upgrade II conditions. Additionally, the DFEI functionality is expected to be expanded, to include neutral reconstructed particles, charm-hadrons decays and particle-identity information. This can bring potential new complementary applications of DFEI, such as providing enhanced flavour-tagging capabilities to LHCb.

Regarding speeding up the inference, a design optimisation of the NP module (for example substituting the GNN by a combination of independent multi-variate classifiers per particle) together with an overall hyperparameter optimisation can bring large reductions to the evaluation time. Significant additional speed ups can be gained by converting the full DFEI pipeline into C++~\cite{Lazar:2022ixi, An_2023} (which is by itself a technical requirement to run DFEI in the current LHCb trigger). The combination of the suggested improvements gives a good hope to achieve the target event-processing rate discussed in Sec.~\ref{sec_timing}. Finally, regarding the utilisation of DFEI in the LHCb Upgrade II, the inference of the GNN modules could become much faster by the usage of GPUs~\cite{Ju:2021ayy,Pata:2022wam} or FPGAs~\cite{Que:2022kmo,Elabd:2021lgo,Heintz:2020soy,Iiyama:2020wap} as hardware accelerators in the trigger system.
\section{Conclusion}
\label{sec_conclusion}

This is the first proof of concept for an inclusive event processing at the LHC in a high-multiplicity environment focused on the identification and explicit reconstruction of all the heavy-hadron decay chains in the event. It is heavily based on deep learning and uses GNNs to optimally capture the event structure. 
To keep the approach computationally scalable, the algorithm is divided into three stages: node pruning removes all the nodes that are not associated with a heavy-hadron decay, edge pruning removes all the edges that do not share the same ancestor inference and finally the lowest-common-ancestor that predicts the hierarchical decay relations of particles, allowing to completely reconstruct all decays. The algorithm has been trained using a simulated dataset that emulates LHCb Run 3 conditions, and is specialised for beauty hadron decays and charged reconstructed particles.

The algorithm is able to separate between particles originating from b-hadron decays and those from the rest of the event better than previous approaches in similar conditions at LHCb. The resulting fraction of perfectly-reconstructed b-hadron decay chains is in the ballpark of the one obtained by the FEI algorithm in an electron-positron environment, showing not only the feasibility but also the competitiveness of this approach at the LHC.

The performance of DFEI is studied in detail both at the global event level and at the individual b-hadron decay level, using both inclusive and exclusive samples containing typical decays of interest for LHCb. A particularly good performance is found in the exclusive modes, in terms of both the efficiency of a perfect decay-chain reconstruction (in the range $20-40\%$) and the efficiency to identify all the reconstructed particles originated from the decay (above $90\%$ in most of the cases).

The application of the algorithm for data analysis at offline level is discussed, explaining how DFEI can be used as a common tool to identify and classify different types of background. These capabilities can already be explored with the Run~3 dataset, which is currently being collected.

In the Upgrade II conditions, the saturation in the rate of events containing b-hadron decays, combined with the large event sizes expected and the limited disk storage capacity, will bring an inverse relation between the amount of information to be stored per event and the total number of events that can be recorded. In terms of charged reconstructed particles, the current DFEI algorithm achieves a $14\times$ event reduction factor in Run 3 conditions, for a $94\%$ efficiency in the selection of particles from b-hadron decays in the event. For illustration, if this kind of performance was achieved in Upgrade II conditions and all the event information was solely related to charged particles, the saving factor would translate into a $14\times$ larger integrated luminosity that could be recorded, compared to storing the full event information. This shows the strong potential of the DFEI approach, while accurate estimates of the gaining factor in Upgrade II conditions will be the focus of future research.

To be used in the trigger, the DFEI algorithm needs to be able to process events at high rate. The speed requirements of the current LHCb trigger system are discussed, a first timing study of the DFEI algorithm is performed, and several steps towards achieving the target event-processing rate are identified.

Finally, the successful development of the DFEI prototype opens the door to future research towards expanding its functionality and use cases in LHCb
and can inspire similar developments in other LHC experiments for the HL-LHC Phase.

\begin{appendices}

\section{Construction of the dataset}
\label{sec_simulation}

To approximately emulate the topology of a Run 3 event in LHCb, single proton-proton collisions at a center-of-mass energy of 13 TeV are first generated with PYTHIA8, using an inclusive \texttt{softQCD} interaction model. Then, several collisions are combined in each event, such that their number follows a poisson distribution with an average of 7.6, corresponding to the average multiplicity expected in Run 3 conditions~\cite{Bediaga:2013tje}. For all the studies done on inclusive b-hadron decays, at least one collision that has produced b-hadrons is included in each event. For the studies done on exclusive decays of interest, the dataset generated in the previous configuration is reused, substituting a collision that produced an inclusive b-hadron decay by a new one, that produced the specified exclusive decay. This new collision is generated by combining the PYTHIA8 and EvtGen generators.

To place those collisions in the space, a coordinate system is defined with its center on the nominal collision point, considered here to be the center of the Vertex Locator of LHCb.
The true position of each proton-proton collision point is sampled from a three-dimensional gaussian distribution, centered on the origin of coordinates, with widths of 0.05 mm, 0.05 mm and 10 mm, respectively along the $x$, $y$ and $z$ axes (see Ref.~\cite{Bediaga:2013tje} for discussions on the expected beam geometry).

Charged stable particles (pions, kaons, protons, electrons and muons) produced in the collisions are only kept if their pseudorapidity is in the range $1.9\leq\eta\leq4.9$, corresponding to the geometric acceptance of LHCb~\cite{LHCb:2008vvz}, and if their origin position along the $z$ direction is within a distance of $\pm500$ mm from the origin of coordinates, which emulates an approximate coverage of the Vertex Locator~\cite{HENNESSY201797}. The measurement of the relevant properties for each particle performed by the LHCb detection and reconstruction process is emulated by modifying the particle properties generated by PYTHIA8, as discussed in the following.

As a first step, the measurement of primary vertices is considered. Those which result in a number of charged particles less than four are considered not to be reconstructible, and hence are discarded. For all the others, a gaussian smearing is applied to their position in each of the three dimensions. The resolution for that smearing as a function of the total number of charged particles in the collision is assumed to be the same as the one measured by LHCb in Run 2 as a function of the number of tracks, that is reported in Fig.~5 of Ref.~\cite{LHCb:2018zdd}. The resolution for the $x$ and $y$ coordinates is assumed to be identical.

In a second step, the determination of the origin point of each particle is studied, which in real life would correspond to the measurement of the position of the first hit from the associated track in the Vertex Locator. The Vertex Locator is segmented into 52 measurement planes along the z-direction~\cite{HENNESSY201797}, which in this study are approximated to be equally spaced for simplicity. The $z$-coordinate of the origin point is therefore assigned to that of the closest plane to the true origin position, looking in the positive direction of the $z$ axis. The values for the $x$ and $y$ coordinates are determined by obtaining the true position of the particle at the given $z$ plane, assuming constant velocity, and applying a gaussian smearing with a resolution of $8.5\;\mu m$ in both the $x$ and $y$ directions (see Ref.~\cite{Billoir:2021srr} for discussions on expected resolutions in Run 3 conditions).

Finally, the measurement of the three-momentum of the particles is emulated. The momentum slope in the $x$ and $y$ directions relative to the $z$ axis is smeared using a gaussian function, using the momentum-dependent resolution reported in Fig.~1 of Ref.~\cite{Billoir:2021srr}. The modulus of the momentum is smeared using a gaussian function assuming a relative resolution of $0.4\%$~\cite{LHCb:2014uqj}.

It should be noted that this emulation does not include additional particles produced in material interactions within the detector or fake particles resulting from wrongly reconstructed tracks, both of which are present in the official LHCb simulation.

\section{Data taking conditions at LHCb Run 3}
\label{sec_detailed_timing_constraints}

In Run 3, the trigger system of LHCb is fully-software based and composed of two consecutive levels, HLT1 and HLT2~\cite{LHCbCollaboration:2014vzo}. The first level performs a partial reconstruction of charged particles, reconstructs primary vertices and performs muon identification, pre-selecting the events to reduce a 30 MHz input rate to 1 MHz~\cite{LHCbCollaboration:2717938}. The raw event information for each passing event is temporarily written to a disk buffer, which allows to perform the next trigger steps in an asynchronous way, and hence with an enlarged computing-time budget. Running on the data in the disk buffer, the HLT2 level performs a full reconstruction of the objects in each event, followed by a combination of inclusive and exclusive selections. Those selections primarily identify interesting events, but can also be used to decide which elements inside them (particles, raw-event information, etc.) will be stored for future processing~\cite{Aaij:2019uij}. The HLT2 output is saved on a permanent tape storage. Before being moved to disk storage, the only one which is accessible for data analysis, the events on the tape undergo a further filtering stage offline~\cite{Skidmore:2022rza}.

The DFEI algorithm can ideally be run at the HLT2 stage, if its inference is fast enough. This would imply event-processing frequencies per computing node in the ballpark of the current HLT2 sequence, which amounts to around 500 Hz excluding selection algorithms~\cite{LHCB-FIGURE-2022-005}. It should be noted that, if this requirement eventually turned out to be too difficult, DFEI could run instead in the Run 3 offline filtering stage before data is sent to disk storage. This would however require persisting the information of all the reconstructed particles to tape.

\section*{Acknowledgements}

J. G. P. has received support from the European Union’s Horizon 2020 research and innovation programme under the Marie Sklodowska-Curie grant agreement No 892683-LHCbDFEI.
A.M. gratefully acknowledges the financial support from the Swiss National Science Foundation (SNF) under project P400P2\_191121. J. E. and N. S. have received support from the Swiss National Science Foundation (SNF) under contract 200020\_204238.
We acknowledge support from the Italian national funding agency INFN. We gratefully acknowledge the provided IT resources of the Hasso Plattner Institut Future Service-Oriented Computing Lab for the research activities, particularly for the usage of GPUs to train the GNN modules.
We would like to thank prof. Enza Messina for fruitful discussions. We would also like to thank the members of the RTA and DPA projects of LHCb for useful comments on the paper.

\section*{Declarations}

\paragraph{Conflict of interest} On behalf of all authors, the corresponding author states that there is no conflict of interest.

\end{appendices}


\bibliographystyle{naturemag}


\end{document}